# Control of excitation transfer in coupled quantum dots by a nonresonant laser pulse

P.A. Golovinski[1,2], V.A. Astapenko[1], A.V. Yakovets[1]

[1] *Department of Radio Engineering and Cybernetics, Moscow Institute of Physics and Technology (State University), Moscow 141700, Russia*

[2] *Physics Research Laboratory, Voronezh State University of Architecture and Civil Engineering, Voronezh 394006, Russia*

We study theoretically fast transfer of excitons between pairs of coupled quantum dots driven by the optical Stark effect that is produced by a short nonresonant laser pulse. The Schrödinger equation, in which the relative position of energy levels of quantum dot subsystems is time-dependent, is solved numerically. Computer simulation shows a way to achieve efficient excitation transfer by the action of a picosecond laser pulse with a rectangular envelope function.

## I. INTRODUCTION

The latest development of supercomputers makes ever greater demands on the electronic components. As a consequence, the attention of researchers is increasingly attracted by coupled quantum dots (QDs) as a basis for a new technology. These "artificial atoms" possess atom-like properties, but they have undeniable advantages over atomic gases since they can provide different energy scales and other physical characteristics that can be easily changed in a wide range of parameters [1]. QDs coupled by the Coulomb interaction [2] are becoming promising candidates for applications in the electro-optical devices and switches, in the same line with systems based on single-electron transfer of electrons [3, 4]. Advantage of such hardware, is manipulation with exciton states in QDs, an opportunity to operate with laser pulse and realize a controlled excitation transfer from one QD to another due to the Förster resonant interaction [5, 6] without a charge motion, accompanied by phenomena of Coulomb blockade and Ohmic dissipation. QDs, consisting of InAs and GaAs, are thought an appropriate for fabrication of all-optical transistor [7].

The theoretical description of such class of phenomena needs further development for better understanding of their detailed mechanism and estimation of the achievable rate of the excitation transfer switching. In QDs, differing in sizes and compositions, excitons with varies energies are observed. It is possible to control the position of the exciton energy levels in QDs by electrostatic or electric field with an optical frequency. In the first case, the DC Stark effect arises [8, 9] that contains an energy shift term for two coupled QDs that is being linear with respect to the field. This effect is a result of a potential difference between the QDs. In the second case, the AC Stark effect is quadratic over the field strength [10, 11]. The use of the optical Stark effect seems most promising in this system since it provides an opportunity of fast control of the relative position of levels in adjacent QDs and, as a consequence, of resonant transfer of energy between them. The shaped pulses can control the excitation dynamics and produce optimum coherence in an atom-like system [12-17]. New possibilities may be opened utilizing THz pulses, for which high polarizability of QDs has been measured experimentally [18]. However, for the present their application for switching exciton transfer is constrained by practical difficulties in formation unipolar pulse.

We consider the exciton states with the lowest energy in two adjacent QDs and an exciton transfer control with the use of picosecond pulses. The barrier between quantum dots is assumed to be high enough to exclude tunnel transition of an electron between the dots and at the same time to be providing considerable Coulomb interaction between excitons of adjacent QDs. It is proposed to use nanophotonics tools, in application of which the minimum size of the area of field localization is not limited by diffraction spot, since interaction can be carried out in the near field of nanosized elements as well as under conditions of confinement and superfocusing of plasmons [19-21]. The coupling between excitons in almost identical quantum dots was observed experimentally in the Ref. 22 dedicated to operation of an all-optical quantum gate. Controlled resonant transfer of excitons was studied theoretically in the Ref. 24 for a system of two semiconductor quantum dots in the presence of the Stark effect [23] in an electrostatic field. The chosen model was based on an approximate Hamiltonian allowing describe the influence of a resonant laser pulse, the Coulomb interaction, the DC Stark effect, and the relaxation of exciton states on system dynamics. With the use of this Hamiltonian, calculations of efficiency of exciton transfer for different conditions of excitation and positions of QD energy levels were carried out. They showed the controllability of a process with the use of a level shift in electrostatic field. The present work is dedicated to construction of a quantum model for the exciton resonant transfer under the action a nonresonant optical pulse, resulting to AC Stark switching.

## II. HAMILTONIAN OF THE SYSTEM AND BASIC EQUATIONS

The main properties of the Förster energy transfer between excitonic states in semiconductors can be analyzed on the basis of a model of a system with two resonant levels [25]. The features of energy transfer depend on the relationship between the magnitude of dipole-dipole interaction energy and the level width [26]. In the case of small distances between quantum dots, the Coulomb interaction dominates [27], energy transfer from a donor, that is an excited QD, to an acceptor proceeds coherently, and a new state is formed, being a superposition of subsystem states. If the acceptor level width is more than the donor level width and the value of interaction, the excitation transfer is carried out incoherently, dissipative processes in the donor proceed slowly, and the dissipation in the acceptor is rapid. This limiting case corresponds to the classical Förster model.

For description the exciton transfer processes, different model Hamiltonians are used. Besides Coulomb interaction between excitons in adjacent QDs, it is advisable to include in the description an energy level shift due to the Stark effect in an external nonresonant optical field and spontaneous decay of excited states due to emission of photons and phonons. Excitation of quantum dots can be done by a femtosecond laser pulse [28], and the energy transfer occurs on a picosecond time scale. This allows time separation of the process of the QD excitation and the process of the excitation transfer. Excitation of a single QD by an ultrashort laser pulse was studied theoretically in Ref. 29, where the coupling of excitons with a phonon bath was taken into account. We will concentrate on the second stage of the problem consisting in controlled excitation transfer. The Förster interaction was taken into account earlier in the work [30] on the basis of a model Hamiltonian, and Coulomb interaction and level widths were added to a model Hamiltonian for two QDs in Ref. 31. In the latter case, equations, written in the density matrix formalism, permit to obtain an exact analytical solution. The analytical solution of the problem of resonant transitions between quantum dots under the action of laser radiation without considering decay is given in Ref. 32. The listed works outline a set of theoretical approaches developed for solution of this class of problems.

Let us consider the energy levels scheme of two excited QDs (the left (L) and right (R) QDs) for formulation quantum model of controlled exciton transfer. We will denote the energy of the system, in which a quasi-stationary exciton state is excited in the L-QD, by $E_L$, and the energy of the system, in which a quasi-stationary exciton state is excited in the R-QD, by $E_R$. We will describe interaction of QDs with a laser pulse by the operator $V(t)$, and the Coulomb interaction between QDs, providing excitation transfer between them, by the operator $V_F$. The spontaneous width of an exciton state is related to the imaginary part of the complex quasi-energy: for the L-dot $\gamma_L = -2\,\text{Im}\,E_L$, and for the R-dot $\gamma_R = -2\,\text{Im}\,E_R$.

The dependence of $V_F$ on the shape of quantum dots, on their size and distance between them was studied for InAs QDs in a GaAs matrix [5]. When conventional optics is used, a field being formed is spatially homogeneous on scales of several tens of nanometers. We propose utilize the near-field optics [33], which provides an opportunity spatially selective excitation one of two closely spaced QDs. Therefore, as an initial state in the problem of energy transfer, the exciton in the L-QD can be taken. Investigation of the excitation transfer between QDs in case of their simultaneous resonant excitation by a laser field, and the control of level detuning by a DC electric field has shown that taking into account biexciton states somewhat changes quantitative results for high levels of laser excitation, but it has no effect on reproduction of main features of dynamics [34], and we will not incorporate such states in our theoretical model.

Let us take the Hamiltonian of the system as the sum of the Hamiltonian describing noninteracting separate QDs, the Coulomb interaction between near-resonant exciton states in adjacent QDs, and the operator of their nonresonant interaction with a laser pulse. Picosecond optical pulses can be considered to be quasi-harmonic, and they can be specified as the product of the harmonic carrier wave of frequency $\omega$ and slowly varying carrier envelope $f(t)$ for the strength of an electric field in a laser pulse. For this case we will consider the carrier wave frequency to be significantly lower than the resonant frequency of quantum dot excitation. In view of this, we will write the operator of interaction of an electromagnetic field with QDs as

$$H_{\text{int}}(t) = hf(t)\cos(\omega t). \qquad (1)$$

Here the operator $h$ acts on spatial variables of a wave function.
We use atomic units in this paper.

Let us take into account the AC Stark shift of levels [35] in QDs in the second order of the perturbation theory for nonresonant interaction $H_{\text{int}}(t)$ [36]. The energy change $\Delta E_j$ for a level *j* caused by this interaction is proportional to the squared strength of the pulse envelope [37]:

$$\Delta E_j = -\frac{1}{4}\alpha_j(\omega) f^2. \qquad (2)$$

The dynamic polarizability $\alpha_j(\omega)$ of a state *j* is given by the known expression for a monochromatic field of frequency $\omega$:

$$\alpha_j(\omega) = 2\sum_m |d_{mj}|^2 \omega_{mj}/(\omega_{mj}^2 - \omega^2). \qquad (3)$$

The values $d_{mj}$ in the Eq. (3) represent dipole matrix elements between the exciton state *j* and some virtual state *m*.

The basis for further description of optical excitation dynamics and controlled energy transfer between QDs is the Schrödinger equation ($\hbar = 1$)

$$i\begin{pmatrix}\dot{\psi}_L \\ \dot{\psi}_R\end{pmatrix} = \tilde{H}(t)\begin{pmatrix}\psi_L \\ \psi_R\end{pmatrix} \qquad (4)$$

with the matrix Hamiltonian

$$\tilde{H}(t) = \begin{pmatrix}\tilde{E}_L(t) & V_F \\ V_F & \tilde{E}_R(t)\end{pmatrix}, \qquad (5)$$

in which $\tilde{E}_L(t) = E_L + \Delta E_L$, $\tilde{E}_R(t) = E_R + \Delta E_R$. When there is no external laser field, $E_L = E - i\gamma_L/2$, $E_R = E - \varepsilon - i\gamma_R/2$. Here $E$ is the real part of the energy of a quasi-stationary state of an exciton in the L-QD, $E - \varepsilon$ is the real part of the energy of a quasi-stationary state of an exciton in the R-QD at the resonance detuning $\varepsilon$. The imaginary parts of the energy $\gamma_L, \gamma_R$ are responsible for exponential relaxation of excitations. We assume that QDs, with similar exciton energy levels, have different polarizabilities (with a difference of values about 10%) due to somewhat different compositions and sizes, so at a certain strength of an external laser field the levels of two QDs coincide due to a corresponding optical Stark shift.

The behavior of the system essentially depends on parameters of a laser field, QDs, and initial conditions, that is, on an initial prepared state. We will consider the action of a short pulse, in which there are no resonant frequencies with QDs, but the system dynamics is changed due to the change of internal resonance conditions, providing controlled efficiency of excitation energy transfer. The specific parameters allowing calculations for excitons in InGaAs/GaAs quantum dots are given in Ref. 30. For example, for quantum dots of $R = 5$ nm parameters take the values: $E = 1.3$ eV, $|\langle \mathbf{r} \rangle| = 6$ Å, $\gamma \cong 10^9 \, s^{-1}$ ($\tau_{decay} = 960$ ps), $V_F = 0.69$ meV ($\tau_F = 0.95$ ps). Such values of parameters mean that the relaxation time is long in comparison with the time of excitation transfer on a picosecond time scale, and, in case of excitation of QDs by a femtosecond laser pulse, the process of controlled transfer can really be considered separate from excitation, assuming existence of completely formed initial exciton state.

## III. ENERGY TRANSFER UNDER THE ACTION OF A NONRESONANT PULSE

For a nonresonant action of the picosecond pulses on the system and considerably long exciton relaxation times, the control of a varying level energy shift during the development of the whole dynamic process reduces to a control of the resonance detuning $\varepsilon$. The Eq. (4) with the Hamiltonian (5) describes direct and reverse coherent resonant transfer of excitation in view of resonance detuning and relaxation of states [38].

An efficiency to control resonant transfer of excitation can be estimated within the framework of a "simple-man model", in which instantaneous switching-on and instantaneous switching-off of a laser pulse providing resonant conditions of transfer for a time $\tau$ are assumed:

$$\varepsilon(t) = \begin{cases} \varepsilon_0, t < 0 \\ 0, 0 < t < \tau \\ \varepsilon_0, \tau < t \end{cases} \quad (6)$$

As initial conditions, we will take $\psi_L(0) = 1$, $\psi_R(0) = 0$. If $\varepsilon = \text{const}$, the solution to the Eqs. (4), (5) can be written as $a_j \exp(-i\varepsilon_k t)$, and the characteristic equation for the system of ordinary differential equations has the form

$$\begin{vmatrix} \tilde{E}_L - \varepsilon_k & V_F \\ V_F & \tilde{E}_R - \varepsilon - \varepsilon_k \end{vmatrix} = 0 \qquad (7)$$

with eigenvalues

$$\varepsilon_{1,2} = \frac{\tilde{E}_L + \tilde{E}_R}{2} \pm \sqrt{\left(\frac{\tilde{E}_L - \tilde{E}_R}{2}\right)^2 + V_F^2} \;. \qquad (8)$$

The general solution to the Eqs. (4), (5) is of form

$$\psi_L = b_L e^{-\gamma_1 t/2} e^{-i\Omega_1 t} + c_L e^{-\gamma_2 t/2} e^{-i\Omega_2 t},$$

$$\psi_R = b_R e^{-\gamma_1 t/2} e^{-i\Omega_1 t} + c_R e^{-\gamma_2 t/2} e^{-i\Omega_2 t}. \qquad (9)$$

where $\Omega_k = \operatorname{Re}\varepsilon_k, \gamma_k = -\operatorname{Im}\varepsilon_k/2$. In view of the initial conditions $b_R = -c_R$, $b_L + c_L = 1$, one have $\dot{\psi}_A(0) = -iV_F$. Hence, it follows that the state of an exciton in the R-QD varies with time according to the equation

$$\psi_R = \frac{V_F}{\Omega_1 - \Omega_2 + i(\gamma_2 - \gamma_1)/2} \left(e^{-\gamma_1 t/2} e^{-i\Omega_1 t} - e^{-\gamma_2 t/2} e^{-i\Omega_2 t}\right), \qquad (10)$$

and the population of the R-state varies by the law

$$|\psi_R(t)|^2 = \frac{V_F^2}{(\Omega_1 - \Omega_2)^2 + (\gamma_2 - \gamma_1)^2/4} \left(e^{-\gamma_1 t} + e^{-\gamma_2 t} - 2e^{-(\gamma_1+\gamma_2)t/2} \cos[(\Omega_2 - \Omega_1)t]\right). \qquad (11)$$

Within the framework of this "simple-man model", the population of the R-QD level increases and decreases periodically, experiencing exponential decay. In the strong coupling limit $V_F \gg \gamma_L, \gamma_R, \varepsilon$, and at $\gamma = \gamma_L = \gamma_R$ population pumping with decay obeys the equation

$$|\psi_A(t)|^2 = e^{-\gamma t}(1 - \cos(2V_F t))/2. \qquad (12)$$

If after the time interval $\tau = \pi/4V_F$ the system is quickly removed from resonance, one will obtain the R-QD population

$$|\psi_R(t)|^2 = \exp(-\pi\gamma/4V_F).  \qquad (13)$$

Such a local maximum for the parameter relation $V_F = 20\gamma$ is 0.93, and the excitation transfer for this case is very high. A real laser pulse that efficiently controls the optical Stark shift of levels should have a near-rectangular "Π"-shape of an envelope, and its switching-on and switching-off should occur for femtoseconds.

For further analysis of excitation transfer under the action of laser pulses with more realistic envelope shape, we will use the numerical solution of the Schrödinger matrix equation, in which, to exclude fast oscillations, a common shift of energy level count was carried out by the replacement $\psi_{L,R}(t) = a_{L,R}\exp(-iEt)$. This replacement imparts the following form to the Schrödinger equation:

$$i\begin{pmatrix} \dot{a}_L \\ \dot{a}_R \end{pmatrix} = \begin{pmatrix} \Delta E_L(t) - i\gamma_L/2 & V_F \\ V_F & \Delta E_R(t) - \varepsilon - i\gamma_R/2 \end{pmatrix} \begin{pmatrix} a_L \\ a_R \end{pmatrix}. \qquad (14)$$

Let us consider how an exciton transfer between QDs depends on a shape of electromagnetic pulses. In fact, a so-called inverse problem arises here, that is, a problem of optimal control of the QD system. A consistent approach to its solution requires finding of a target functional maximum under some constraints [39]. We will significantly simplified the problem by using the property proved in the Ref. 40 on the basis of the Pontryagin's maximum principle, this property consisting in the fact that optimal control of quantum systems with a discrete spectrum is realized by instantaneous switching-on resonant field. It is obvious that in this case the envelope functions are preferable that make it possible to achieve the most sharp and fast switching-on and switching-off of a pulse, that is, fast rising to the maximum value of field amplitude providing the fulfilment the exact resonance condition.

Let us consider several kinds of pulse carrier envelopes with different shapes. The first kind is based on representation of a pulse as the product of arctangents. For this kind, the time dependence of detuning energy in view of the variable time-dependent Stark shift has the form

$$\varepsilon(t) = \varepsilon_0 \left(1 - \left(\arctan\left(\frac{t-t_1}{\tau}\right) + \frac{\pi}{2}\right)^2 \left(\arctan\left(\frac{t_2-t}{\tau}\right) + \frac{\pi}{2}\right)^2 / \pi^4 \right), \tag{15}$$

where $\varepsilon_0$ is the initial resonance detuning. The choice of parameters $t_1, t_2$, and $\tau$ makes it possible to change both instants of switching-on and switching-off of a pulse and the duration of irradiation mode change. The form of the dependence specified by the Eq. (15) is shown in Fig. 1.

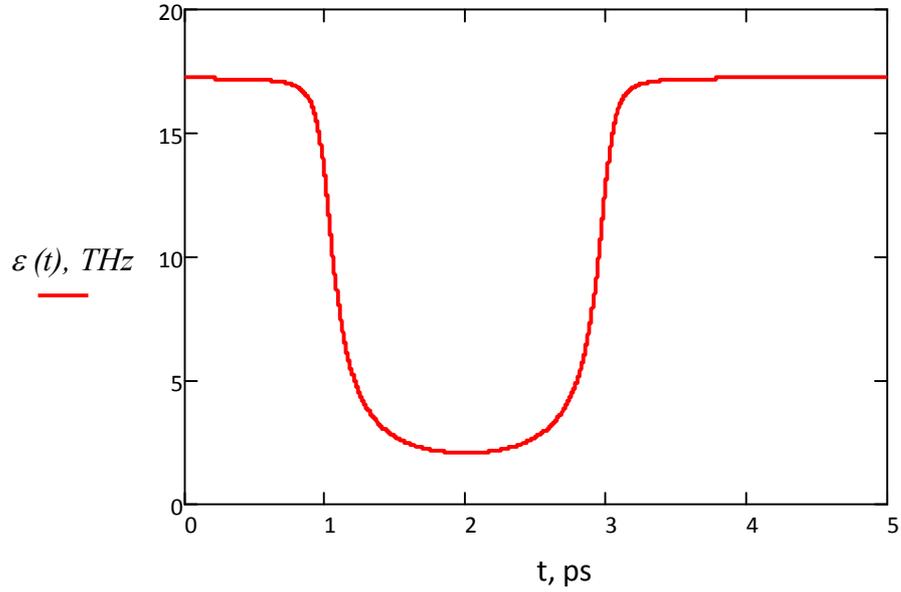

FIG. 1. The dependence of the Stark energy shift expressed in terms of the time function of the arctangents product. Switching-on occurs at an instant of time of 1 ps, switching-off occurs at an instant of time of 3 ps, $\tau = 0.1$ ps.

The second pulse carrier envelope shape is described by the standard Gaussian function, and

$$\varepsilon(t) = \varepsilon_0 \left(1 - \exp\left(-(t-t_1)^2 / \tau^2\right)\right) \tag{16}$$

Within the framework of such a functional dependence, one cannot control the time of switching-on and switching-off of a field, since $t_1$ is responsible only for the time of achievement of the maximum value of the envelope function, and $\tau$ is responsible for its width.

The third kind pulse is based on the experimental results [41] for generation of rectangular laser pulses. Initial pulses in the authors' experimental setup were spectrally limited Gaussian optical pulses generated with the use of passive mode locking of a tunable laser and duration from 600 fs to 1.8 ps at a central wavelength of 1535 nm. The principle of generation is based on the fact that every desired ultrafast temporal waveform can be synthesized as a superposition of a Gaussian pulse and its successive time derivatives since they form bases of wavelets [42]. Flat-top intensity waveform pulses are well approximated by two terms of the general series, that is, by a proper combination of a Gaussian pulse and its first time derivative. As an optical differentiator, a homogeneous long-period fiber lattice was used [43]. The technology has made it possible to synthesize flat-top waveforms of different durations in the subpicosecond mode. A remarkable property of the proposed method of generation rectangular optical pulses is the fact that rectangular pulses of different durations can be synthesized with supply a Gaussian pulse of proper duration to a linear system. The time dependence of intensity for experimental and two theoretical pulse waveforms is shown in Fig. 2.

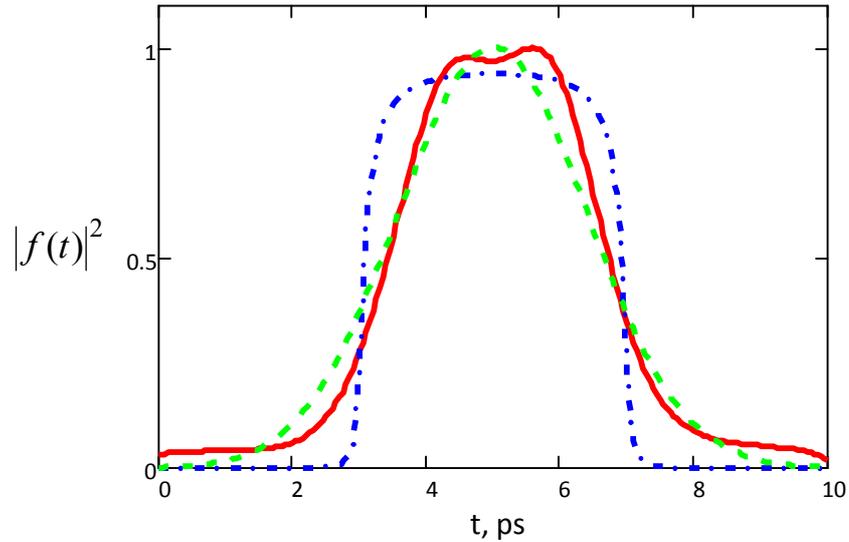

FIG. 2. The time dependence of optical pulse intensity. The blue dash-and-dot line represents the temporal profile of pulse intensity specified analytically with the use of the product of arctangents, the green dotted line shows a pulse expressed with the exponent, the red solid line reproduces an experimental pulse.

The estimation of QD polarizability, taking into account two levels in the Eq. (3), the QD parameters from Ref. 30, and the pulse parameters from Ref. 41, gives the value $\alpha(\omega) \approx 400 \cdot 10^{-24}$ cm$^3$. Then with a difference of polarizabilities of L- and R-QDs of 10% the strength of an electric field of a laser pulse at a maximum should reach the value $f = 25 \cdot 10^5$ V/cm.

It is of interest to compare how the time dependence of the probability of the system be in the states $|L\rangle$ and $|R\rangle$ looks in case of switching-on electromagnetic pulse specified by the product of arctangents and the Gaussian carrier envelope. Shown in Fig. 3 is the result of calculation of the excitation transfer with level quasi-crossing in the controlled Stark effect for a pulse of the shape given by the Eq. (15). An electromagnetic pulse was switched on at an instant of time of 1 ps and switched off at an instant of time of 3 ps. The maximum population of the acceptor level is 0.93, and it is reached for 2.5 ps. As follows from the results of calculation, the excitation transferred from the donor state to the acceptor not so efficiently as in the case a perfect rectangular pulse, but there is no significant reverse process of energy transfer back to the donor.

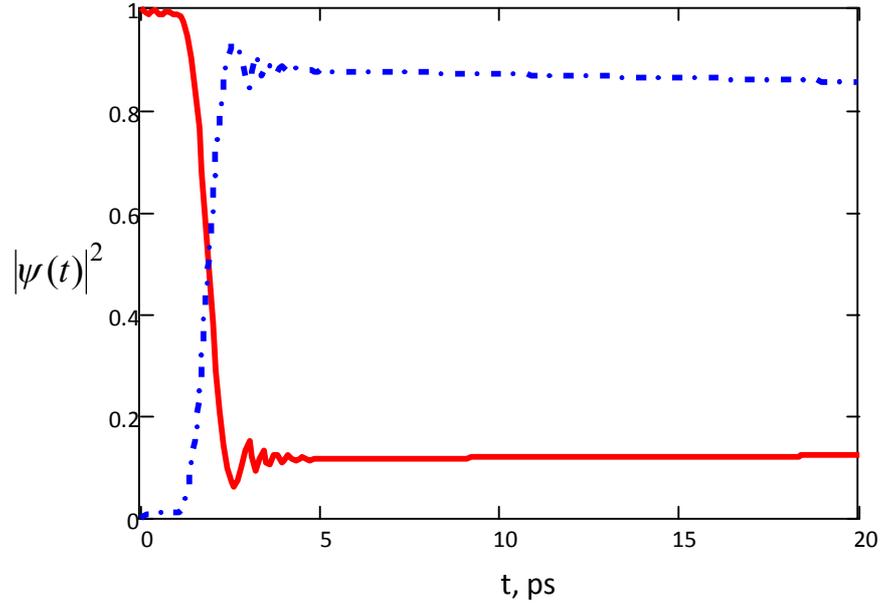

FIG. 3. The time dependence of the probability of the system $|\psi_L|^2$ be in the state $|L\rangle$ (red curve) and of the probability of the system $|\psi_R|^2$ be in the state $|R\rangle$ (blue dash-and-dot curve). The shape of a picosecond electromagnetic pulse is taken as the product of arctangents.

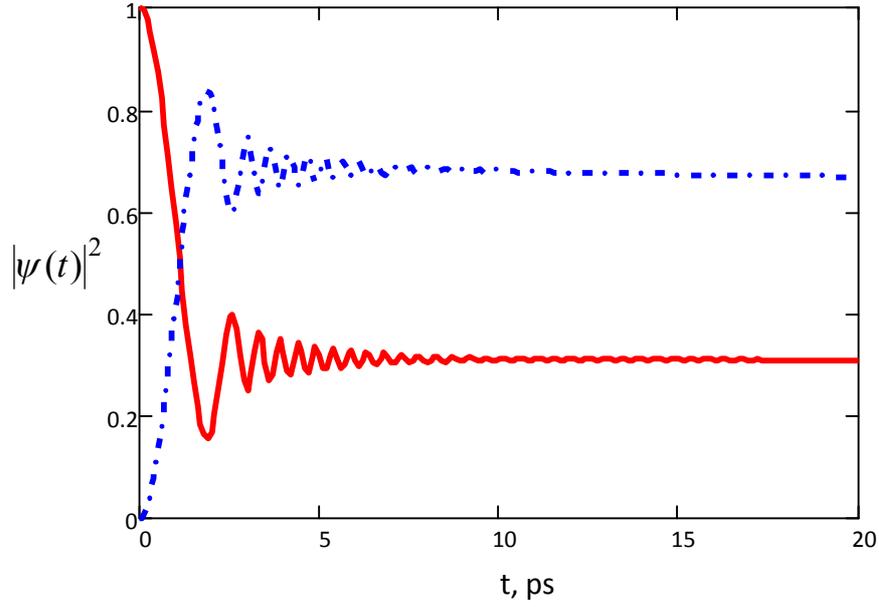

FIG. 4. The time dependence of the probability of the system be in the state $|L\rangle$ (red curve) and in the state $|R\rangle$ (blue dash-and-dot curve). A picosecond electromagnetic pulse has a Gaussian shape.

For comparison, the control of the excitation transfer has been calculated with the use of the Gaussian pulse envelope, for which it is impossible to achieve sharp switching-on and switching-off a field during femtosecond time interval. Shown in Fig. 4 is the best result of calculation of excitation transfer with level quasi-crossing in the controlled Stark effect in this case. An electromagnetic pulse reaches the maximum value at $t_1 = 1$ ps. The maximum population of the acceptor level is 0.85, and then it decreases to 0.7. Small oscillations show multiple, partially reversible energy transfer between the donor and the acceptor. As seen from the data in the Fig. 4, the excitation transferred from the donor state to the acceptor not so efficiently as in case of the exact rectangular envelope function or the arctangents product. The efficiency of excitation transfer decreased since the pulse has a smooth envelope shape, and the system stays in the resonance range too little time to complete transfer.

To approach the experimental parameters, we will use one of the time dependences of pulses obtained in Ref. 41, and apply it for description of optical control of the excitation transfer between QDs.

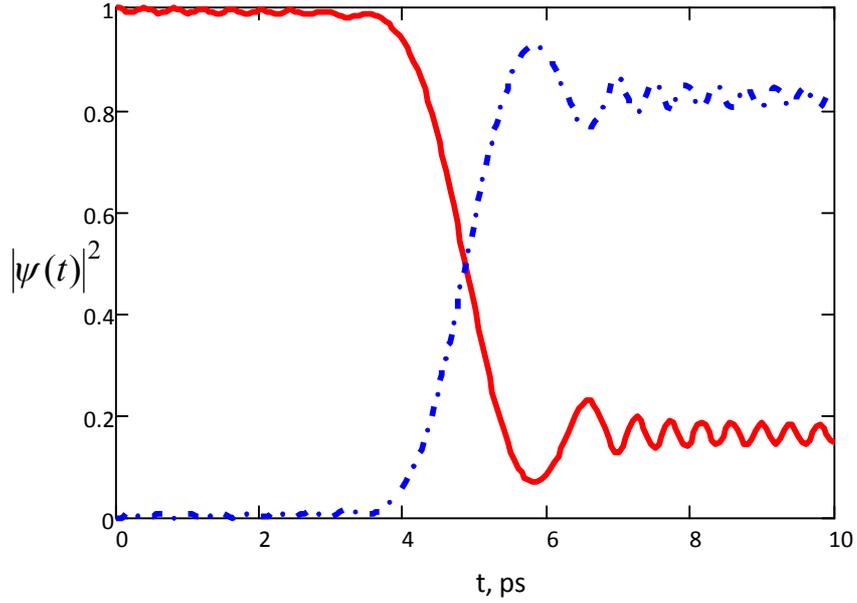

FIG. 5. The time dependence of the probability of the system being in the state $|L\rangle$ (red curve) and in the state $|R\rangle$ (blue dash-and-dot curve). The experimental pulse envelope shape is used.

Shown in Fig. 5 is the result of calculation of excitation transfer with level quasi-crossing in the controlled Stark effect under the action of a flat-top pulse. The maximum population of the acceptor level is 0.93, and then after termination of a pulse it decreases to 0.8. Small oscillations show multiple, partially reversible energy transfer between the donor and the acceptor. When the experimental pulse waveform is used in computer simulation calculations, as seen from the data given in Fig. 5, optical control of excitation transfer proceeds even more efficiently than with the use of the theoretical model (15) of flat-top waveforms.

**IV. SUMMARY**

We have studied theoretically quantum dynamics of exciton transfer between QDs in control of the nonresonant optical Stark effect under the action of a short laser pulse. A model Hamiltonian was formulated that allows the description of laser control of the energy transfer between QDs. The analytical estimation of efficiency the energy transfer between QDs in a two-level model was given, and efficient numerical methods of calculation of such processes in view of nonresonant optical control and relaxation were developed on the basis of solution of the Schrödinger equation in the matrix form. Carried out computer simulation shows the reproduction of main physical phenomena in systems of two QDs connected with coherent energy exchange and decay due to relaxation processes.

The notion of energy transfer acquires real significance only on condition that the size of a system is great enough in comparison with the spatial region of excitation, and if the initial state is formed by a femtosecond laser pulse with the use of a near-field optics. Otherwise, both QDs can be excited simultaneously if the Rabi frequency overlaps the energy distance between levels of the excitons. In this case direct excitation of the whole system, being under consideration, by optical photons can produce directed energy transfer only as a result of selective relaxation of excitation in a dedicated QD. The developed model approximations allow calculate the excitation energy transfer between various QDs in a wide range of change of QDs system parameters and optical excitation. The proposed scheme shows opportunity of efficient control of the exciton transfer between QDs by variation a level shift due to the AC Stark effect and a promising outlook for further experimental investigation of QDs as components of all-optical nanotransistors.

**ACKNOWLEDGEMENTS**

The research was supported by the Russian Foundation for Basic Research (grant No. 13-07-00270) and by the Government order of the RF Ministry of Education and Science (project No. 1940).